\documentclass[twocolumn,pre,showpacs]{revtex4}
\usepackage{epsfig}
\usepackage{graphicx}
\usepackage{epsfig}
\usepackage{amsmath}
\usepackage{amssymb}
\usepackage{latexsym}
\usepackage{aeguill}
\usepackage{dcolumn}
\usepackage{bm}
\begin{document}
%

\title{On the absence of the glass transition in two dimensional hard disks}
\author{Marco Tarzia$^{a,b}$}
\affiliation{${}^a$ Dipartimento di Scienze Fisiche and INFN sezione di Napoli,
Universit\`{a} degli Studi di Napoli ``Federico II'',
Complesso Universitario di Monte Sant'Angelo, via Cinthia, 80126 Napoli, Italy}
\affiliation{${}^b$ School of Physics and Astronomy, University of Manchester, 
Manchester, M13 9PL, United Kingdom}
\date{\today}

\begin{abstract}
In this paper we study the glass transition in a model of identical hard 
spheres, focusing on the two dimensional case.
In the mean-field limit the model exhibits an ideal glass
transition of the same nature of that found in discontinuous spin glasses.
Nevertheless, a systematic expansion around the mean-field solution 
seems to indicate that the glass transitions is smeared out in two dimensions,
in agreement with some recent results.
Our investigation could be generalized to higher spatial dimensions, 
providing a way to determine the lower critical dimensionality
of the mean-field ideal glass picture.
\end{abstract}
\maketitle

Under fast enough cooling or densification diverse materials, such as molecular
and polymeric liquids, colloidal suspensions, granular assemblies, molten
mixtures of metallic atoms, may form glasses~\cite{debenedetti}, 
i.e., amorphous states that may be 
characterized mechanically as a solid, but 
lack of the long range crystalline order.
Despite all the work devoted to the subject,
the underlining mechanisms responsible for the vitrification processes 
are not well understood, as the transition to the glassy state 
is still deemed specifically to be one of the most obscure enigmas in
condensed matter physics.
Many valuable theories attempt to describe these remarkable phenomena,
but none of them is as yet regarded as compelling.

A system of identical hard 
spheres confined in a fixed 
volume~\cite{hardspheres,hardglass,biroli,zamponi2d,torquato,krauth,zamp_larged,zamp_hnc} 
is the simplest system exhibiting a dramatic slowing down of the dynamics
in the high volume fraction region, referred
by many authors as a glass 
transition~\cite{hardglass,biroli,zamp_larged,zamp_hnc}.
In a recent paper~\cite{zamp_larged}, using a replica
approach it has been very nicely 
shown that hard spheres in large space dimensions
undergo an ideal glass transition at a volume fraction $\phi_K$. 
The same result has been obtained in three dimensions employing 
diverse kinds of mean-field-like approximations (such as the 
hypernetted chain approximation~\cite{zamp_hnc} and the small cage 
expansion~\cite{zamp_larged,zamponi2d}) and using the Carnhan-Starling 
equation of state.
However, for finite dimensional systems (with short range interactions)
the mean-field picture should be modified by 
(non-perturbative) activated events~\cite{activated}, 
and one might wonder to what extent the
mean-field scenario is still valid in that case.

Here we study a system of identical hard disks on a two dimensional square 
lattice.
We first analyze the mean-field solution, where 
an ideal glass transition of the same nature of that 
found in mean-field model for glasses~\cite{pspin} 
occurs. We then consider a systematic
expansion around the mean-field limit, which allows to
take into account short range correlations as corrections to
the mean-field approximation. This is accomplished by considering 
bigger two dimensional $L \times L$ square cells of size $L = 2, 3, \ldots$, 
within which the model is solved
exactly (see also Ref.~\cite{CAM,rizzo}). 
We observe that the glass transition occurs at higher
densities as the size of the cell 
is increased, and seems to be smeared out in the
limit $L \to \infty$. This analysis hints that there is no 
glass transition in two dimensional hard disks, in agreement with the 
recent results of Refs.~\cite{zamponi2d,torquato,krauth}. 
Interestingly enough, the study presented here could be easily generalized
to higher spatial dimensions, providing a direct 
way to estimate the lower critical
dimensionality of the mean-field ideal glass picture. 
The latter investigation could prune
down the number of candidate theories for the glass transition.

The Hamiltonian of the model reads:
\begin{equation}
{\cal H} = \sum_{[i,j]} J \, n_i n_j,
\end{equation}
where the lattice variables $n_i = 0,1$ whether the cell $i$ is occupied 
by a particle or not, and the sum is restricted over the couples of
sites $[i,j]$ such that their distance is equal or less than two lattice 
spacings: $d_{i,j} \le 2$. The model can 
be regarded as a system of hard disks of diameter $\sqrt{5}-\epsilon$ 
lattice spacings.
The limit $J \to \infty$ is taken, insuring the hard core exclusion.

\begin{figure}
\begin{center}
\psfig{figure=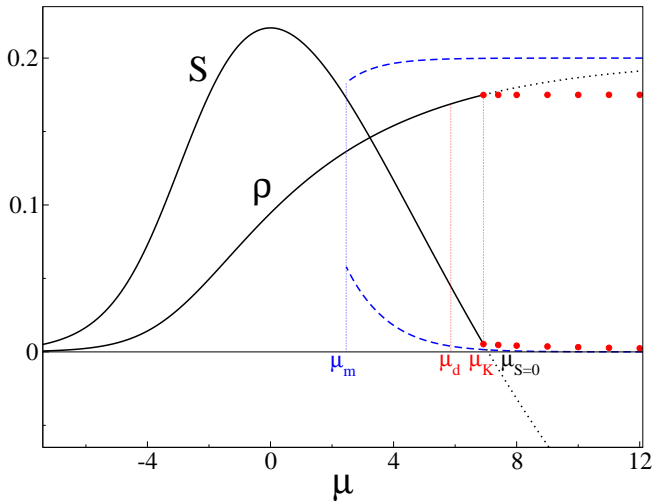,scale=0.31,angle=270}
\end{center}
\caption{Mean field solution of the  
model on the random graph. Density, 
$\rho$, and entropy per site, $S$, as a function of the chemical potential,
$\mu$, in the liquid (black continuous curve), crystalline (blue dashed curve), 
and glassy (red circles) phases.}
\label{fig:meanfield}
\end{figure}

The model can be solved in mean-field on the random regular 
graph~\cite{lgm}, i.e., 
a random lattice where every vertex has $k+1$ neighbors,
but which is otherwise random (to mimic the 
two dimensional case, we choose $k=3$). Locally the graph has a
tree-like structure with a finite branching ratio, but has loops of
typical size $\ln N$. The presence of loops is crucial to insure the 
geometric frustration, but the local tree structure allows for an analytical
solution of the model, since we can write down iterative equations on the
local probability measure. To this aim, let us consider a branch of the
tree ending on the site $i$, 
and denote by $i_j$, $j \in \{ 1 , 
\ldots , k \}$, the roots of the branches connected to the site $i$. 
We call
$Z_1^{(i)}$ the partition function of this branch 
restricted to the configurations where the site $i$ is occupied by
a particle. Analogously, we define $Z_0^{(i)}$  
the partition function of the branch restricted 
to configurations where the site $i$ is empty, and $\overline{Z}_0^{(i)}$
the partition function of the branch restricted
to configurations where the site $i$ is empty with all neighbors $i_j$ empty.
Using the Grand Canonical ensemble, 
the following recursion relations are derived:
\begin{eqnarray} \label{eq:iterz}
\nonumber
&& Z_1^{(i)} = e^{\beta \mu} \prod_{j=1}^{k} \overline{Z}_0^{(i_j)}, \qquad \,\,
\overline{Z}_0^{(i)} = \prod_{j=1}^{k} Z_0^{(i_j)},\\
&& Z_0^{(i)} = \left[ 1 + \sum_{j=1}^{k} \frac{Z_1^{(i_j)}}{Z_0^{(i_j)}}
\right] \prod_{j=1}^{k} Z_0^{(i_j)},
\end{eqnarray}
where $\mu$ is the chemical potential. 
It is convenient to introduce on any site $i$ the local cavity fields
$\beta h_i = \ln ( Z_1^{(i)} / Z_0^{(i)})$ and
$\beta v_i = \ln ( \overline{Z}_0^{(i)} / Z_0^{(i)})$, in term
of which the iteration relations, Eqs.~(\ref{eq:iterz}), read:
\begin{eqnarray} \label{eq:iterf}
\nonumber
e^{\beta h_i} &=& e^{\beta \mu + \sum_{j=1}^{k}  
\beta v_{i_j} } \bigg (
1 + \sum_{j=1}^{k} e^{\beta h_{i_j}} \bigg)^{-1},\\
e^{\beta v_i} &=& \bigg ( 1 + \sum_{j=1}^{k} e^{\beta h_{i_j}} \bigg)^{-1}.
\end{eqnarray}
From these fields one can obtain the free energy, $\beta F = - \ln Z$, as
a sum of site and link contributions~\cite{lgm}: $F = \Delta F_s -
(k+1) \Delta F_l /2$. The contribution from the bond between
two branches with root sites $i_1$ and $i_2$ is:
\begin{equation} \label{eq:dfl}
e^{- \beta \Delta F_l} = 1 + e^{\beta(h_{i_1} + v_{i_2})} + 
e^{\beta(h_{i_2} + v_{i_1})},
\end{equation}
while the contribution from the addition of a site $i$ connected with
$k+1$ branches with root sites $i_j$ reads:
\begin{equation} \label{eq:dfs}
e^{- \beta \Delta F_s} = 1 + \sum_{j=1}^{k+1} 
e^{\beta h_{i_j}} + e^{\beta \mu + \sum_{i=1}^{k+1} \beta v_{i_j} }.
\end{equation}
Starting from Eqs.~(\ref{eq:iterf}), we find at low density a liquid phase, 
characterized by a homogeneous (replica symmetric) solution,
$h_i = h$ and $v_i = v$. Given this solution, using Eqs.~(\ref{eq:dfl}) and
(\ref{eq:dfs}), the thermodynamic quantities can be derived, and, in
particular, the density $\rho = \langle n_i \rangle$ 
and the entropy per lattice site, $S = 
- \beta F - \beta \mu \rho$, 
are obtained (continuous curve in Fig.~\ref{fig:meanfield}). 
As the chemical potential (the density) is
increased, a first order phase transition from the liquid phase to
a crystalline phase occurs at a melting point $\mu_m \simeq 2.46$ ($\rho_m
\simeq 0.1364$). The crystalline state is characterized by a periodic
structure, which breaks down the translational invariance, and can be
obtained introducing different sub-lattices.
More precisely, we introduce three sub-lattices, $a$, $b$, and $c$, on 
which the local cavity fields are site independent. The sub-lattices
must be organized in such a way to reproduce the crystalline order:
each vertex of the sub-lattice $a$ is connected with $3$ sites of the 
sub-lattice $b$ [i.e., $a \to (b,b,b)$]. Analogously, we have $b \to (c,c,c)$
and $c \to (c,c,a)$. Also the free energy shifts, Eqs.~(\ref{eq:dfl}) and
(\ref{eq:dfs}), must be computed carefully, taking into account the 
structure of the three sub-lattices.
In the crystalline phase, as the chemical potential is 
further increased above $\mu_m$, the density  
rapidly approaches the maximum density, $\rho_{max}
= 0.2$, and the entropy per site approaches zero (dashed curves in 
Fig.~\ref{fig:meanfield}).

The crystallization transition can be avoided and, in this case, the system
enters a supercooled state, still described by the homogeneous solution of  
Eqs.~(\ref{eq:iterf}). 
However, as shown in Fig.~\ref{fig:meanfield}, 
the entropy per site in the supercooled liquid becomes
negative when the density is increased above $\rho_{S=0} \simeq 0.1757$ 
(or the chemical potential is increased above $\mu_{S=0} \simeq 7.07$). As a 
consequence, a thermodynamic phase transition must occur at a density
$\rho \lesssim 
\rho_{S=0}$. In fact, in the mean-field approximation we find that the
system undergoes a phase transition toward a 1RSB glassy phase, which
can be analyzed by taking into account 
the existence of many different local minima (or pure states) 
of the free energy. Since in this
case the local fields can fluctuate from pure state to pure state, this 
situation is described by a probability distribution $P (h,v)$ that
the fields $h_i$ and $v_i$ on the site $i$ equal $h$ and $v$ for a 
randomly chosen pure state. Using the cavity method~\cite{lgm} we find that
this function satisfies the self-consistent equation:
\begin{equation} \label{eq:1rsb}
\frac{P(h,v)}{{\cal N} e^{- \beta m v}}  
=  \int  \prod_{j=1}^{k} [ \textrm{d} h_{i_j} 
\textrm{d} v_{i_j} P(h_{i_j}, v_{i_j}) ] \delta (h - h_i)
\delta (v - v_i),
\end{equation}
where ${\cal N}$ is a normalization constant, $h_i$ and $v_i$ are the
local cavity fields obtained when merging $k$ branches which carry
the fields $(h_{i_j}, v_{i_j})$ [via Eq.~(\ref{eq:iterf})], and $m \in [0,1]$
is a Lagrange multiplier which turns out to be the usual 1RSB 
parameter, fixed by the maximization of the free energy
with respect to it~\cite{lgm,SGT}.

\begin{figure}
\begin{center}
\psfig{figure=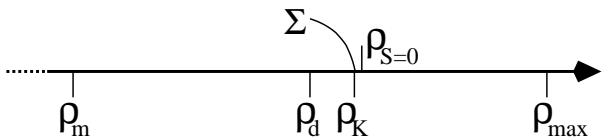,scale=0.31}
\end{center}
\caption{
Schematic drawing of the location of the relevant densities
emerging from the mean-field solution: melting density,
$\rho_m \simeq 0.1364$, dynamical transition density, $\rho_d \simeq
0.1688$, Kauzmann transition density, $\rho_K \simeq 0.1750$, 
density at which the entropy of the supercooled liquid vanishes, $\rho_{S=0}
\simeq 0.1757 (\sim \rho_K)$, 
maximum density of the crystalline packing,
$\rho_{max} = 0.2$. The figure also show that the configurational entropy, 
$\Sigma$, defined as the logarithm of the number of pure states, 
is positive for $\rho_d \le \rho \le \rho_K$.}
\label{fig:dens}
\end{figure}

The 1RSB cavity equation, Eq.~(\ref{eq:1rsb}), can be solved exactly in
the close packing limit ($\beta \to \infty$),   
where the recursion relations, Eq.~(\ref{eq:iterf}), simplify to:
\begin{equation}
h_i = 1 + \sum_j v_{i_j} - V(h_{i_j}), \qquad
v_i = - V(h_{i_j})
\end{equation}
with $V(h_{i_j}) = \textrm{max} \{ h_{i_j} \} \theta (\textrm{max} 
\{ h_{i_j} \})$  [we have set $\mu = 1$]. These equations yield an exact
ansatz for the cavity fields probability distribution: 
$P(h,v) = \sum_{r=1}^7 
p_r \delta(h - h_r) \delta(v - v_r)$, with $h_{1,2,3,4} = 1,
0, -1, -2$, $v_{1,2,3,4} = 0$, $h_{5,6,7} = 0, -1, -2$, and $v_{5,6,7} = -1$.
By defining $P = p_1$, $R = p_2 + p_3 + p_4$, and $Q = p_5 + p_6 + p_7 = 1 - 
P - R$, Eq.~(\ref{eq:1rsb}) becomes:
\begin{eqnarray}
P & = & R^3 \, \left[ e^y - (e^y - 1)(1 - P)^3 \right]^{-1},\\
\nonumber
R & = & \left[ (1 - P)^3  - R^3 \right] \, \left[ e^y - (e^y - 1)(1 - P)^3 
\right]^{-1},
\end{eqnarray}
where $y = \lim_{\beta \to \infty} \beta m$.
In terms of $P$ and $R$, the 1RSB link and site contribution to the 
free energy read:
\begin{eqnarray}
\Delta \phi_l [y] &=& - y^{-1} \ln \left \{ 1 + (e^y - 1)P^2 
+ 2 e^y P R \right \}\\
\nonumber
\Delta \phi_s [y] &=& - y^{-1} \ln \left \{ e^y + (e^y - 1) \left[
R^4 - (1 - P)^4 \right] \right \}.
\end{eqnarray}
The free energy is then given by $\phi [y] = \Delta \phi_s [y]
- 2 \phi_l [y]$, from which we can compute the complexity $\Sigma
 = y^2 \partial \phi [y] / \partial y$ and the density $\rho = 
 \partial (y \phi[y]) / \partial y$.
 
The finite $\mu$ solution of the 1RSB cavity equations can be
found numerically, using the population dynamics algorithm~\cite{lgm}. 
In agreement with the results of Refs.~\cite{zamp_larged,zamp_hnc}, the 
mean-field solution of the model exhibits an ideal glass transition of the same 
nature of that found in mean-field discontinuous spin glasses~\cite{pspin}.
We first find 
a purely dynamical transition at $\rho_d \simeq 0.1688$
($\mu_d \simeq 5.86$), where a non 
trivial solution of Eq.~(\ref{eq:1rsb}) appears for the
first time,
signaling the emergence of an extensive number of
metastable states (which, in mean-field, trap the dynamics for an infinite
time). A solution of the 1RSB equation becomes thermodynamically stable
at a higher density, $\rho_K \simeq 0.1750$ ($\mu_K \simeq 6.92$), 
where a thermodynamic
transition to a 1RSB glassy phase takes place.
The relevant densities emerging from the mean-field approximation are
reported in Fig.~\ref{fig:dens}, showing 
that $\rho_K$ is strikingly close to $\rho_{S=0}$. 

\begin{figure}
\begin{center}
\psfig{figure=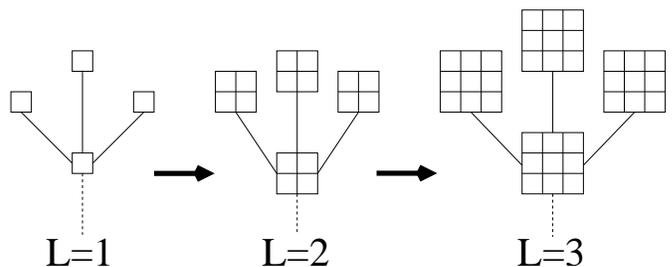,scale=0.293}
\end{center}
\caption{Expansion around the mean-field solution. Starting from the
mean-field limit ($L = 1$), bigger finite-dimensional
cells of size $L=2$ and $L=3$ are considered. 
}
\label{fig:cells}
\end{figure}

In the following we consider a systematic expansion around the mean-field 
theory, which takes into account the actual structure of the two dimensional
square lattice. More precisely, we consider cells of size $L=2,3$ within
which the model is solved ``exactly'' and we use those cells as vertex
of the mean-field theory on the random graph, as depicted
in Fig.~\ref{fig:cells} (see also Refs.~\cite{CAM,rizzo}).
In the limit $L \to \infty$ the exact solution of the model should be
achieved.
This method allows to include in an exact fashion short range
spatial correlations, as correction to the mean-field limit. Since in glassy
systems there is no diverging {\em equilibrium} length scale, this expansion
is expected to be reliable and effective.

\begin{figure}
\begin{center}
\psfig{figure=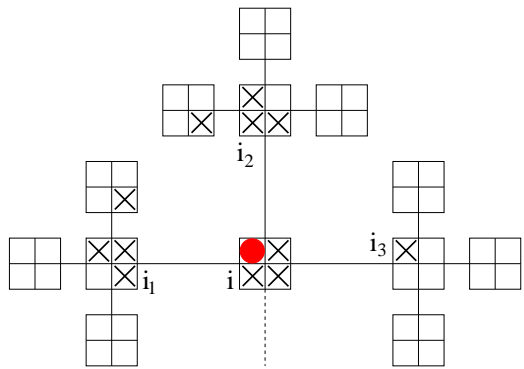,scale=0.34}
\end{center}
\caption{
Case $L=2$:
the figure depicts a configuration in which a particle (red circle) occupies 
one of the two forward sites of the branch ending on the cell $i$, i.e., one
of the two sites which are not on the edge of the cell where the link
is missing.
The sites whose distance from the particle is less or equal than two
lattice spacings cannot be occupied (these sites are marked by crosses).}
\label{fig:two}
\end{figure}

In order to solve the $L=2$ case,  
let us consider a branch of the tree ending on the $2 \times 2$
square cell $i$,
and denote by $i_j$, $j \in \{ 1 ,
\ldots , k \}$, the roots of the branches connected to the cell $i$,
as shown in Fig.~\ref{fig:two}.
We call $U^{(i)}$ (resp., $D^{(i)}$) 
the partition function of the branch restricted to 
the configurations in which the cell is occupied by a particle in one 
of the two ``forward'' (resp., ``backward'') sites, i.e., one of the two 
sites which {\em are not} on the edge of the cell where the link is missing 
(resp., one of the two
sites which {\em are} on the edge of the cell where the link is missing).
We also define $\overline{U}^{(i)}$ as the partition function of the 
branch restricted to the configurations in which the cell is occupied by 
a particle in one of the two forward sites, with its ``left''
(or, equivalently, ``right'') neighbor cell constrained to be not occupied
in one of the two backward sites.
Finally, we need to introduce $Z_0^{(i)}$, defined as the partition
function of the branch restricted to configurations in which the cell
is empty, and $\overline{Z_0}^{(i)}$, corresponding to the partition
function of the branch restricted to configurations in which the cell
is empty with its ``left''
(or, equivalently, ``right'') neighbor cell constrained to be not occupied
in one of the two backward sites.
Introducing four local cavity fields, defined as
$\beta u_i = \ln ( U^{(i)} / Z_0^{(i)})$, 
$\beta d_i = \ln ( D^{(i)} / Z_0^{(i)})$,
$\beta v_i = \ln ( \overline{U}^{(i)} / Z_0^{(i)})$,
and $\beta p_i = \ln ( \overline{Z_0}^{(i)} / Z_0^{(i)})$,
within the RS homogeneous ansatz for the supercooled liquid, the 
following algebraic recursion relations are derived:
\begin{eqnarray}
\nonumber
e^{\beta u} &=& e^{\beta \mu} T^2 W Z^{-1}, \qquad
e^{\beta v} = e^{\beta \mu} T^2 Y Z^{-1}\\
e^{\beta d} &=& e^{\beta \mu} T \left( e^{\beta d} + W \right) Y Z^{-1}\\
\nonumber
e^{\beta p} &=& \left [ e^{\beta d} \left ( W e^{\beta d} + Y \cdot F \right )
 + Y \left( e^{\beta d} + F \right) W \right] Z^{-1},
\end{eqnarray}
where $T = e^{\beta p} + e^{\beta v}$, $Y = 1 + 2 e^{\beta u}$,
$W = 1 + e^{\beta d} + 2 e^{\beta u}$, $F = W + e^{\beta d}$, and
$Z = \left [ e^{\beta d} \left ( W e^{\beta d} + Y \cdot F \right )
 + W^2 \left( e^{\beta d} + F \right) \right]$.
In terms of the local cavity fields, the 
contribution to the free energy from the addition of a link
between two cells, $\Delta F_l$, and from the addition of a cell, $\Delta F_s$,
can be computed:
\begin{eqnarray}
e^{-\beta \Delta F_l} & = & Y^2 + 4 e^{\beta d} \left( e^{\beta v} + 
e^{\beta p} \right)\\
\nonumber
e^{-\beta \Delta F_s} & = & Y \left[ 
4 e^{\beta \mu} T^2 \left( W + e^{\beta d} \right) + Z \right] 
+ 2 e^{\beta (d + p)},
\end{eqnarray}
from which one can derive the free energy per site 
$F = (\Delta F_s - 2 \Delta F_l)/4$, the density of particles
$\rho = \langle n_i \rangle = 
- ( \partial F / \partial \mu ) /4$, and the entropy per site
$S = -\beta F -\beta \mu \rho$.

\begin{figure}
\begin{center}
\psfig{figure=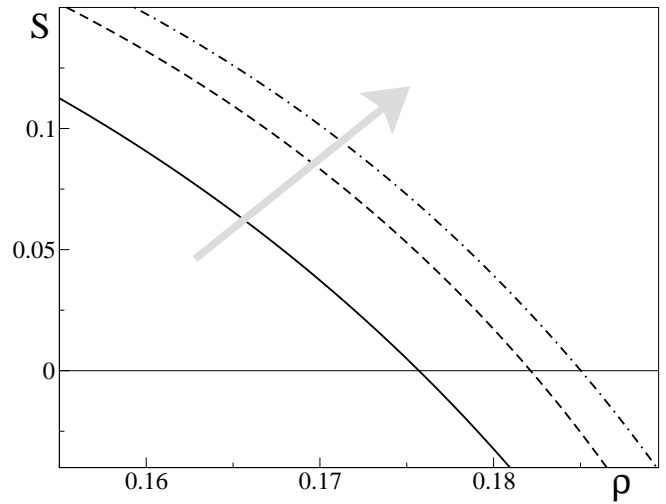,scale=0.31,angle=270}
\end{center}
\caption{Entropy per site, $S$, as a function of the density, $\rho$.
L increases along the direction of the arrow:
$L=1$ (continuous curve), $L=2$ (dashed curve)
and $L=3$ (dotted-dashed curve). The density at which the entropy of the
supercooled liquid vanishes, $\rho_{S=0}$, increases as
$L$ is increased.}
\label{fig:entropy}
\end{figure}

The recursion relations for the case $L=3$ can be determined using a
similar procedure.
In Fig.~\ref{fig:entropy},
the entropy, $S$, is plotted as a function of the density, 
$\rho$, for $L=1$, $L=2$ and $L=3$, in the homogeneous solution for the 
supercooled liquid. The figure shows that the expansion around the
mean-field limit systematically modifies the results. In particular, 
we note that the density at which the entropy of the supercooled liquid 
vanishes, $\rho_{S=0} (\sim \rho_K)$, moves toward higher densities
when $L$ is increased. As a consequence, 
the instability of the supercooled liquid
(and, therefore, the transition to the glassy phase) 
is displaced toward the maximum density when bigger two dimensional
cells are considered. 
In fact, we find that 
$\rho_{S=0}$ approaches very nicely $\rho_{max} = 0.2$ as a 
power law: $\rho_{S=0} (L) \simeq 0.2 - 0.024 \, L^{-0.44}$. 

Further insights can be gained by studying the value of the entropy of
the supercooled liquid in the zero temperature limit, defined as 
$S_{\infty} = \lim_{\mu \to \infty} S$.
$S_{\infty}$ is negative in the mean-field approximation [$S_{\infty} 
(L=1) \simeq - 0.291$] and increases systematically as $L$ is increased.
We find that 
$S_{\infty}$ nicely approaches zero as a power law as a function of $L$:
$S_{\infty} \simeq - 0.291 \, L^{-0.36}$.


These results clearly hints that the instability of the supercooled liquid 
and the thermodynamic transition to the glassy phase is smeared out in
two dimensions, as one includes corrections to the mean-field theory: 
the entropy of the supercooled liquid seems to vanish only at the maximum
density of the crystalline state.
This is in agreement with the recent findings of 
Refs.~\cite{zamponi2d,torquato}.
Our results are also in agreement with those of Ref.~\cite{krauth},
where, by employing a suitable Monte Carlo algorithm, the authors show that
there is no evidence for a thermodynamic phase transition up to 
very high densities in two dimensional (polidisperse) hard disks; 
the glass is thus indistinguishable from the liquid on 
purely thermodynamic grounds.
Note, however, that in Ref.~\cite{reich} the authors state that 
numerical claims in favour of and/or against a thermodynamic glass 
transition must be considered carefully, due to the difficulties 
to probe the system close enough to $\rho_K$.

If the liquid has to be a good solution in the $L \to \infty$
limit up to $\rho_{max}$, 
its pressure must diverge at this point. 
In fact, we find 
that for every value of $L$, the pressure of the supercooled liquid
diverges for $\mu \to \infty$, i.e., the entropy of the liquid
approaches $S_{\infty}$ at $\rho_{max}$ with a vertical slope.
Since, as discussed above, 
$S_{\infty}$ is extrapolated to zero and $\rho_{S=0}$ is extrapolated
to $\rho_{max}$ as $L$ is increased, it seems reasonable to expect that 
in the large $L$ limit the entropy of the liquid vanishes at $\rho_{max}$
and that, consistently, the pressure diverges at this point.

However, it is important to highlight that 
this is not a {\em proof} but just a {\em hint} 
of the absence of the glass
transition in hard disks: as a matter of fact, 
$\rho_{S=0}$ is only an upper bound to $\rho_K$. Thus
it might be possible that, in the limit of large $L$, $\rho_K < \rho_{max}$ 
even if $\rho_{S=0} \to \rho_{max}$. In principle, one should compute
$\rho_K$ for different values of $L$, by solving the 1RSB equations, which is,
unfortunately, a hard numerical task.
Nevertheless, given the closeness of $\rho_K$ to $\rho_{S=0}$ for $L=1$, 
and the consistency of the liquid solution when extrapolated for
large $L$ up to $\rho_{max}$,
one might guess that the possibility described above is unlikely and
that $\rho_{S=0}$ provides a good estimation for $\rho_K$
also for bigger $L$.

In conclusion, we have presented an analytical study of a system of
identical hard spheres, focusing on the case of hard disks on a square
lattice. The mean-field version on the model 
exhibits an ideal glass transition of the same kind of that found
in mean-field discontinuous spin glasses~\cite{zamp_larged}. 
Nevertheless, by considering
a systematic expansion around the mean-field solution able to take into
account short range correlations in an exact fashion, we have shown that
such glass transition seems to be smeared out in two dimensions, 
confirming the results of Refs.~\cite{zamponi2d,torquato,krauth}. 
Note that the results
presented here are also in agreement with the recently discovered mapping 
of glass forming system to Ising spin glasses in an external magnetic 
field~\cite{mike}, according to which there should not be a thermodynamic 
glass transition in dimensions less than six.
Since there cannot be a dynamical glass transition without a thermodynamical
one (provided that the dynamics satisfies the detailed balance), we 
finally argue that there is no structural arrest in two dimensional hard
disks at a density smaller than the one of the crystalline packing.
This analysis could be generalized to the three dimensional case (which is
the most relevant for supercooled liquids) and to higher spatial
dimensions, providing a direct way to investigate the lower critical
dimensionality of the mean-field ideal glass scenario.
This study could prune down the number of candidate theories
for the glass transition.

~

I would like to warmly thank G. Biroli, A. de Candia, 
A. Fierro, P. McClarty, M. A. Moore, and F. Zamponi for useful remarks and
comments.
I would also like to thank A. Coniglio for discussions and for his 
continuous support.
Financial support by
the European Community's Human Potential Programme under contracts
HPRN-CT-2002-00307, DYGLAGEMEM, and  MRTN-CT-2003-504712, ARRESTED
MATTER, is also acknowledged.
Work supported by  
MIUR-PRIN 2004, MIUR-FIRB 2001.

\end{document}